\documentclass[conference]{IEEEtran}
\IEEEoverridecommandlockouts

\usepackage[compress]{cite}
\usepackage{amsmath,amssymb,amsfonts}
\usepackage{algorithmic}
\usepackage{graphicx}
\usepackage{textcomp}
\usepackage{xcolor}
\usepackage[hidelinks]{hyperref}
\usepackage{subcaption}
\usepackage{braket}
\usepackage{float}
\usepackage{xcolor} 

\definecolor{customgreen}{RGB}{45, 180, 85}

\def\BibTeX{{\rm B\kern-.05em{\sc i\kern-.025em b}\kern-.08em
    T\kern-.1667em\lower.7ex\hbox{E}\kern-.125emX}}

\begin{document}

    \title{Solving Large-Scale Vehicle Routing Problems with Hybrid Quantum-Classical Decomposition}

\author{
\IEEEauthorblockN{
Andrew Maciejunes\IEEEauthorrefmark{1}, 
John Stenger\IEEEauthorrefmark{2}, 
Daniel Gunlycke\IEEEauthorrefmark{2}, 
Nikos Chrisochoides\IEEEauthorrefmark{1}\IEEEauthorrefmark{3}
}
\IEEEauthorblockA{
\IEEEauthorrefmark{1}Department of Physics, Old Dominion University, Norfolk, VA, USA \\
\IEEEauthorrefmark{2}Chemistry Division, Naval Research Laboratory, Washington, D.C., USA \\
\IEEEauthorrefmark{3}Department of Computer Science, Old Dominion University, Norfolk, VA, USA
}
}

\maketitle
\IEEEpubid{\vspace{1.2cm}\footnotesize DISTRIBUTION STATEMENT A: APPROVED FOR PUBLIC RELEASE: DISTRIBUTION IS UNLIMITED.}

\begin{abstract}
    We present a two-level decomposition strategy for solving the Vehicle Routing Problem (VRP) using the Quantum Approximate Optimization Algorithm. 
    A Problem-Level Decomposition partitions a 13-node (156-qubit) VRP into smaller Traveling Salesman Problem (TSP) instances. 
    Each TSP is then further cut via Circuit-Level Decomposition, enabling execution on near-term quantum devices. 
    Our approach achieves up to 95\% reductions in the circuit depth, 96\% reduction in the number of qubits and a 99.5\% reduction in the number of 2-qubit gates. We demonstrate this hybrid algorithm on the standard edge encoding of the VRP as well as a novel amplitude encoding.
    These results demonstrate the feasibility of solving VRPs previously too complex for quantum simulators and provide early evidence of potential quantum utility.

\end{abstract}


\section{Introduction}

The Vehicle Routing Problem (VRP) is a well-known NP-hard combinatorial optimization problem that has been studied extensively over the past five decades \cite{widuch2020}. 
The VRP addresses the question: how does one route a given number of vehicles across a set of destinations most efficiently? 
This challenge is highly relevant to large companies such as Amazon and ExxonMobil \cite{ibmExonMobile}. 
The global supply chain and logistics industry is projected to be worth over 13 trillion USD by 2026 \cite{Statista}, prompting a strong demand for scalable 
and efficient routing solutions.

Numerous variants of the VRP exist, each introducing additional constraints or objectives. 
These include the VRP with Time Windows (VRP-TW), the Capacitated VRP (CVRP), and the Open VRP (OVRP). 
Comprehensive surveys of these variants can be found in \cite{kumar2012survey, elatar2023vehicle, elshaer2020taxonomic, irnich2014chapter, vidal2020concise, miller1960integer}. 
The most basic form of the problem is the single vehicle case, commonly referred to as the Traveling Salesman Problem (TSP) \cite{robinson1949hamiltonian,miller1960integer}.
\IEEEpubidadjcol
 
The TSP seeks the shortest path that visits each location exactly once and returns to the starting point. 
The list of possible solutions corresponds to all permutations of the destinations. Table~\ref{tab:permutations_example} shows the complete list of permutations for three cities, A, B, and C. The total number of routes is $3! = 6$.
Assuming symmetric travel costs (i.e., the cost of $A \rightarrow B \rightarrow C$ is equal to $C \rightarrow B \rightarrow A$), the number of unique 
routes can be reduced to:
\begin{equation}
    n_p = \frac{n!}{2}
\label{eq:perumtations_for_tsp}
\end{equation}
where $n$ is the number of cities and $n_p$ is the number of unique permutations.
 Any brute-force solution to this problem has time complexity $\mathcal{O}(n!)$.

Although relatively small values of $n$ are computationally manageable, such as the $n = 3$ case with 3! = 6 permutations; larger instances quickly become intractable.
For example, a problem with $n = 1000$ cities has approximately $10^{2500}$ possible routes. 
Even for a small town, the United States Postal Service may need to deliver daily mail to 1000 homes, making brute-force solutions infeasible. 
The complexity of the VRP only increases as more vehicles are introduced.

\begin{table}[ht]
\centering
\begin{tabular}{|c|c|}
    \hline
    Index & Permutation \\ \hline
    1 & ABC \\ \hline
    2 & ACB \\ \hline
    3 & BAC \\ \hline
    4 & BCA \\ \hline
    5 & CAB \\ \hline
    6 & CBA \\ \hline
\end{tabular}
\caption{All permutations of three cities, A, B, and C.}
\label{tab:permutations_example} 
\end{table}

\begin{figure*}[ht]
    \centering
    \includegraphics[width=\linewidth]{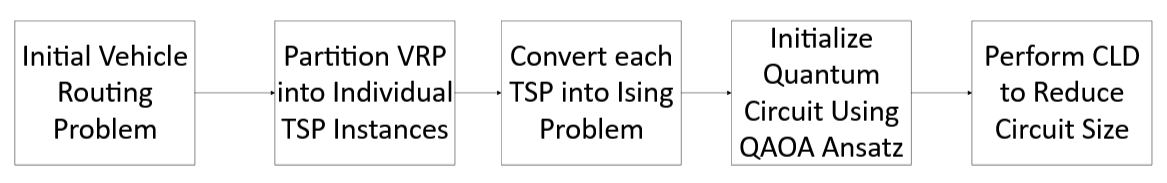}
    \caption{High-level overview of two-level decomposition}
    \label{fig:methodology_flowchart}
\end{figure*}

Due to the factorial growth in solution space shown in Equation~\ref{eq:perumtations_for_tsp}, heuristic-based methods are often employed. 
These algorithms do not guarantee optimality but can produce high-quality approximate solutions in reasonable time. 
The choice of heuristic often depends on the specific variant of the VRP under consideration.

Perhaps the simplest heuristic is the nearest neighbor method. Starting from an initial node, it repeatedly visits the closest unvisited node. 
This greedy algorithm is easy to implement but can yield poor results for certain problem instances \cite{jungnickel1999greedy, gutin2002traveling}. 
More sophisticated approaches include metaheuristics, which incorporate adaptive strategies to better explore the solution space and avoid local minima. 
Reviews of classical algorithms for various VRP variants can be found in 
\cite{liu2023heuristics, konstantakopoulos2022vehicle, li2007open, tan2001heuristic, zhang2022review, elshaer2020taxonomic}.
\IEEEpubidadjcol

Quantum computing has the potential to outperform classical heuristics for certain large problem instances. 
While this advantage remains theoretical, early work has shown success in applying quantum techniques to small-scale VRP cases, such as the 4-node single-vehicle or 
3-node two-vehicle problems \cite{fitzek2024applying}. 
Grover’s algorithm \cite{grover1996fast} offers a quadratic speedup for unstructured search and has been adapted to VRP scenarios \cite{sato2025two}. 
Other quantum algorithms explored in this context include the Quantum Adiabatic Algorithm 
\cite{arino2023adiabatic, kieu2019travelling, farhi2001quantum}, Quantum Phase Estimation \cite{tszyunsi2023quantum, srinivasan2018efficient}, and quantum 
annealing via D-Wave systems \cite{martovnak2004quantum, borowski2020new}.
The Quantum Approximate Optimization Algorithm (QAOA) offers a gate-based alternative to annealing that can be executed on universal quantum computers \cite{farhi2014quantum}. However, these methods are constrained by the limitations of current Noisy Intermediate-Scale Quantum (NISQ) devices.

NISQ devices are characterized by a small number of qubits and significant gate errors~\cite{Preskill2018}.  All modern quantum computers fall into this category.  In the future, error correcting techniques~\cite{Knill1998,Kitaev1997,Shor1995} are expected to be able to provide error-free quantum computing.  However, these techniques require the use of many physical qubits to form one error-corrected logical qubit.  Thus, for the foreseeable future, quantum computers will be limited in the number of available qubits.  Therefore, it is critical to develop quantum algorithms that decompose large problems into smaller problems that can be solved on separate quantum computers.       

In this paper, we present a two-level decomposition strategy to extend the capabilities of quantum hardware, consisting of:
\begin{itemize}
    \item \textbf{Problem-Level Decomposition (PLD)}: Decomposing the VRP into multiple smaller and simpler TSPs.
    \item \textbf{Circuit-Level Decomposition (CLD)}: Decomposing each TSP circuit for execution on NISQ-era quantum hardware.
\end{itemize}
First, we perform PLD, where the VRP graph is divided into multiple smaller graphs, each representing a single TSP instance. 
Each TSP is then mapped to a quantum circuit. To reduce quantum circuit depth and qubit requirements, we apply CLD 
using the methods described in \cite{tang2021cutqc}. This enables execution on currently available Quantum Processing Units (QPUs), which we simulate. 
We evaluate the performance of our two-level decomposition by comparing the total solution cost to that obtained via Google OR-Tools. 
Additionally, we verify the correctness of each TSP instance using brute-force enumeration of all possible permutations.

\noindent \newline
{\bf Contributions} Key contributions are three:

\begin{itemize}
    \item Problem decomposition of VRP into smaller-sized TSP problems
    \item Circuit decomposition and new encoding for the TSP, resulting in up to 95\% reductions in circuit depth and 95 \% CX in 2 qubit gate count and 96\% reduction in qubit count. 
    \item Demonstration of utility-scale quantum computing, for 13-node VRP requiring 156  qubits, with a comparable (and sometimes better) solution to classical VRP methods. Readily available QPUs from IBM contain 133 qubits.
    \item Proof of concept on both QUBO\cite{zahedinejad2017combinatorial} and amplitude \cite{stenger2024efficient} encodings.
\end{itemize}

\section{Methodology}

The methodology is divided into five sections. 
We begin by defining and discussing the VRP. 
We then move to PLD, where we describe the graph partitioning process used to reduce the VRP to 
multiple instances of the TSP. 
Next, we define the cost function and describe the encodings used to map each TSP instance to a quantum device. We implement both an edge encoding and an amplitude encoding.
The fourth section outlines our implementation of QAOA. 
Finally, we conclude with the CLD process used to enable execution of the circuits on near-term quantum hardware. 
Figure~\ref{fig:methodology_flowchart} highlights the overall methodology.

\subsection{Defining the VRP}

\begin{figure*}[t]
    \centering

    \begin{subfigure}[b]{0.48\textwidth}
        \centering
        \includegraphics[width=\linewidth]{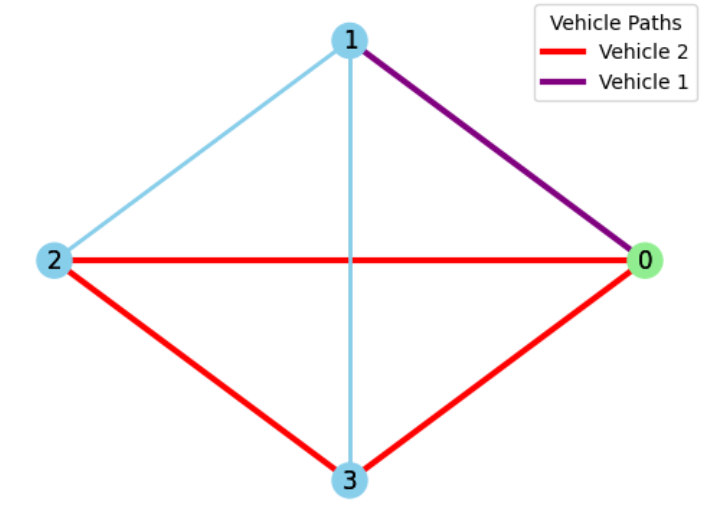}
        \caption{Valid VRP Solution}
        \label{fig:example_vrp_solution}
    \end{subfigure}
    \hfill
    \begin{subfigure}[b]{0.48\textwidth}
        \centering
        \includegraphics[width=\linewidth]{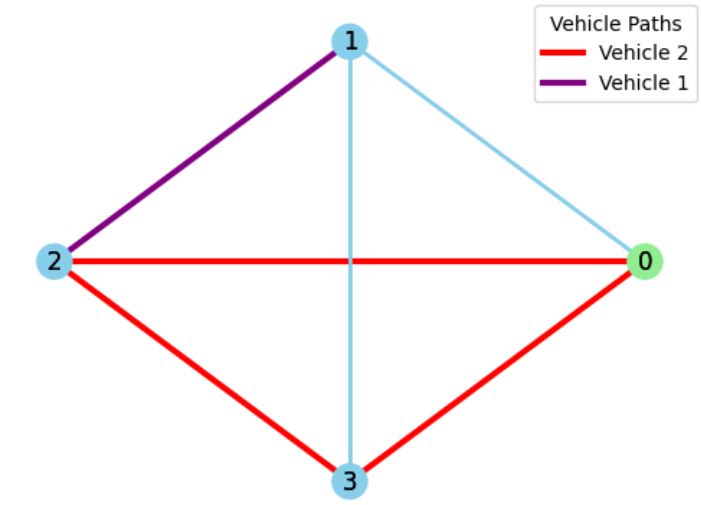}
        \caption{Invalid VRP Solution}
        \label{fig:vrp_example_invalid_solution}
    \end{subfigure}

    \caption{(a) shows a VRP instance of $n = 4$ and $K = 2$. Vehicle 1 (purple) follows the path $0 \rightarrow 1 \rightarrow 0$. Vehicle 2 (red) follows the path $0 
    \rightarrow 2 \rightarrow 3 \rightarrow 0$. Node 0, highlighted green, represents the depot. It should be noted that this path is equivalent to  $0 \rightarrow 3 \rightarrow 2 
    \rightarrow 0$ due to the symmetry of the problem. (b) illustrates a VRP instance of $n = 4$ and $K = 2$. The solution violates both constraints. 
    Vehicle 1 (purple) does not start and end at the depot, and node 2 is visited more than once. The blue lines represent connections between two nodes. These are potential paths from one 
    node to the next.}
    \label{fig:combined_vrp}
\end{figure*}

Formally, the  basic VRP problem description is broken down into 
two statements, (I) Given, and (II) Task. Here is each, according to \cite{VehicleRoutingTextBook}.  

\textit{Given}: A set of transportation requests and a fleet of vehicles. This is traditionally represented as a graph, which is how we will choose to represent it throughout this paper.

\textit{Task}: Determine a set of vehicle routes to perform all (or some) transportation
requests with the given vehicle fleet at minimum cost; in particular, decide
which vehicle handles which requests and the order they are processed so that all vehicle routes
can be optimally executed.

We can represent the VRP as a fully connected, undirected, weighted graph, $G$, with $n$ vertices and $n(n-1)$ edges. Figure~\ref{fig:example_vrp_solution} shows an example. 
Mathematically, we can write it as a mixed integer programming problem with an objective function to minimize and constraints to enforce. We can write the objective function $C$, also referred to as the cost function, as

\begin{equation}
    C(x)= \sum_{i \sim j} w_{ij} x_{ij} \text{ } \forall i\text{, }j \in \{1,...,n\},
    \label{eq:objective_function}
\end{equation}
where $x_{ij}$ represents the edge connecting node $i$ and node $j$,  $w_{ij}$ is the corresponding edge weight, and the summation over $i \sim j$ is a summation over all edges of the graph.
The binary variable, $x_{ij}$, takes value 1 if the edge is in the route, and 
value 0 otherwise. If the edge is in the route, it is referred to as active or activated. The objective function is the sum of edge weights of all activated edges. The goal of the VRP is to find the lowest possible value for the cost function;
\begin{equation}
    C_s(x) = \min_{x_{ij} \in \{0,1\}}C(x)
\end{equation}
where $C_s(x)$ corresponds to the most efficient route.

The two constraints are represented in a similar way. 
The first constraint is that each non-depot node must be visited exactly once. 
This means that each node has exactly one incoming and one outgoing edge. 
The equation

\begin{equation}
    \sum_{j \in \sigma(i)^+} x_{ij} = 1 \quad \forall i \in \{1, \dots, n\},
    \label{eq:constraint_1a}
\end{equation}
where $\sigma(i)^+$ represents the set of all edges going out of node $i$, ensures each node has exactly one outgoing edge. 
We represent exactly one incoming edge as

\begin{equation}
    \sum_{j \in \sigma(i)^-} x_{ji} = 1 \quad \forall i \in \{1, \dots, n\},
    \label{eq:constraint_1b}
\end{equation}
where $\sigma(i)^-$ is the set of all edges coming into node $i$.

The second constraint is that each vehicle must start and end at the depot. 
Mathematically, we enforce this by requiring that \( K \) edges leave the depot and \( K \) edges return to it. 
Specifically, we write

\begin{equation}
    \sum_{i \in \sigma(0)^+} x_{0i} = K ,
    \label{eq:constraint_2a}
\end{equation}
to ensure that each of the \( K \) vehicles departs from the depot, and

\begin{equation}
    \sum_{j \in \sigma(0)^-} x_{j0} = K,
    \label{eq:constraint_2b}
\end{equation}
to ensure that each vehicle returns to the depot. 
The above objective function and constraint equations represent the Miller-Tucker-Zemlin (MTZ) formulation discussed in \cite{miller1960integer}. 
\subsection{Graph Partitioning}

\begin{figure*}[t]
    \centering

    \begin{subfigure}[b]{0.3\textwidth}
        \centering
        \includegraphics[width=\linewidth]{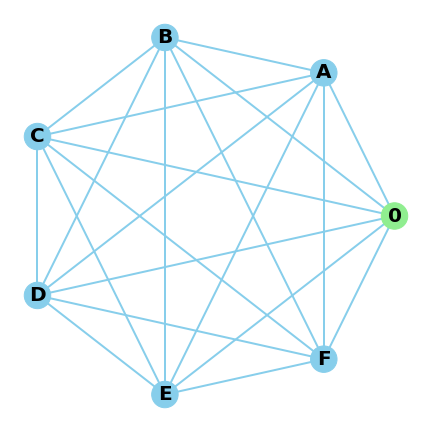}
        \caption{7 Node VRP}
    \end{subfigure}
    \hfill
    \begin{subfigure}[b]{0.3\textwidth}
        \centering
        \includegraphics[width=\linewidth]{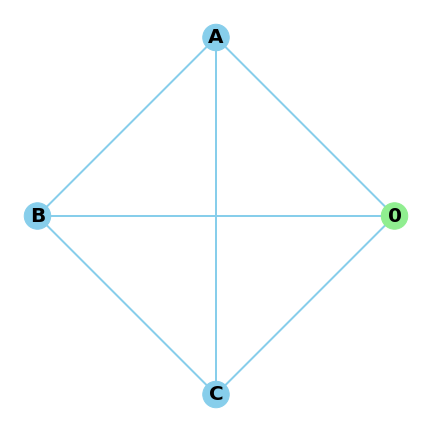}
        \caption{First Decomposition}
    \end{subfigure}
    \hfill
    \begin{subfigure}[b]{0.3\textwidth}
        \centering
        \includegraphics[width=\linewidth]{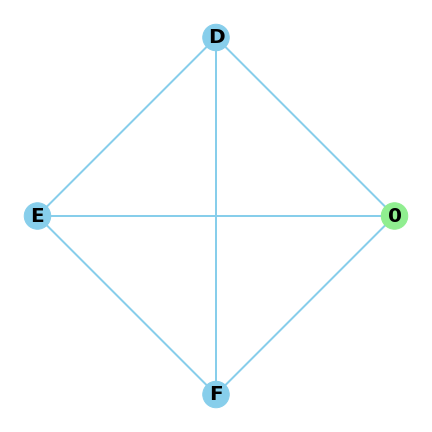}
        \caption{Second Decomposition}
    \end{subfigure}

    \caption{Graph decomposition for a 7-node VRP instance with 2 vehicles. 
    (a) The full graph with 7 cities, where the green node represents the depot requires 42 qubits. The graph will be decomposed into two subgraphs.
    (b) The first subgraph includes the depot and three cities (A, B, C), forming a single TSP instance. This requires 12 qubits.
    (c) The second subgraph includes the depot and the remaining cities (D, E, F), forming another TSP instance. This requires 12 qubits.
    This decomposition reduces the quantum resources required for solving the VRP from 42 qubits once to 12 qubits twice. }
    \label{fig:graph_decomposition}
\end{figure*}

To decompose the VRP into smaller subproblems compatible with near-term quantum hardware, we apply graph partitioning techniques. 
As shown in Figure~\ref{fig:graph_decomposition}, the VRP graph consists of \( n \) nodes—one representing the depot and the rest representing destinations. 
Before partitioning, we temporarily remove the depot node, which is later added back into each partition to maintain the correct VRP structure. 
The number of partitions is equal to the number of vehicles \( K \).  Each resulting subgraph forms a single-vehicle TSP instance.

The primary motivation for decomposition is to reduce the circuit width required to solve each subproblem. 
The full VRP graph often leads to quantum circuits that exceed the available qubit count on current QPUs, denoted \( \xi_{\text{max}} \). 
By partitioning the graph, we aim to ensure that each TSP subproblem has circuit width \( N_q^{(p)} < \xi_{\text{max}} \), enabling execution on NISQ devices. 
This partitioning strategy allows us to distribute quantum resources across subcircuits while maintaining problem fidelity.

For this work, we directly employ the METIS graph partitioning library~\cite{METIS}. 
METIS seeks to minimize the number of edges cut between partitions, using a three-stage process—coarsening, partitioning, and refinement. 
However, using METIS in this context presents several challenges. 
First, METIS does not guarantee that the resulting partitions will be simply connected~\cite{chrisochoides1989automatic}, which is essential for 
solving each partition as a valid TSP. 
A disconnected subgraph may imply an invalid route or an unsolvable subproblem; we implement only fully connected VRPs to prevent this problem.

Second, METIS and similar traditional graph partitioners treat the graph as static, unaware of the downstream quantum constraints such as circuit 
depth and qubit fidelity.
Accordingly, we plan to develop a problem-specific graph partitioner for the Quantum VRP (QVRP) that directly respects hardware limitations by (1) ensuring 
simply connected partitions to minimize reconstruction complexity, and (2) balancing qubit count and gate count per subcircuit to maximize hardware utilization and fidelity.

\subsection{Mapping to a Quantum Computer}

Quantum computers behave differently from classical computers, allowing them to solve certain optimization problems significantly faster \cite{chicano2025combinatorial}. 

Quantum computers work by encoding the problem as a quantum state and leveraging interference, superposition, and entanglement to guide the search toward the optimal solution. 
The problem is ultimately converted into an energy minimization problem, often through methods such as QUBO or directly as an Ising Hamiltonian 
\cite{zahedinejad2017combinatorial}. 
Other encoding techniques exist to map the VRP to a quantum computer, such as Higher Order Binary Optimization (HOBO) \cite{goldsmith2024beyond}, 
which allows for encoding more information into the cost function with fewer binary variables. 
Another approach is the permutation encoding \cite{glos2022space}, which maps permutations to a binary string. 
This encoding requires $\log_2{n!}$ qubits. 
A more qubit-efficient approach is to represent the edges as basis states, as shown in \cite{stenger2024efficient}, which requires $\log_2{(n^2 - n)}$ qubits. 
Despite differences in implementation, all of these approaches ultimately reduce to an energy minimization, or Ising, problem. 
We use the standard QUBO mapping from \cite{zahedinejad2017combinatorial} which requires one qubit for each edge, or 
\begin{equation}
    N_q = (n^2-n)
    \label{eq:num_qubits}
\end{equation}
qubits. 

Using the QUBO method, we create a cost function from the MTZ formulation shown by Equations~\ref{eq:objective_function} through~\ref{eq:constraint_2b}, 
by converting the constraints into penalty terms. 
The idea is that if a constraint is violated, a significant amount of energy is added so that when minimizing the cost function, 
the optimizer avoids any constraint violating states. 
The resulting QUBO is shown as 

\begin{align}
    \mathcal{H}_c(x) &= \sum_{i \sim j} w_{ij}x_{ij} + \lambda \sum_{i \in \{1, \dots, n\}} \left( \sum_{j \in \sigma(i)^+} x_{ij} - 1 \right)^2 \notag \\
    &\quad + \lambda \sum_{i \in \{1, \dots, n\}} \left( \sum_{j \in \sigma(i)^-} x_{ji} - 1 \right)^2 \notag \\
    &\quad + \lambda \left( \sum_{i \in \sigma(0)^+} x_{0i} - K \right)^2 + \lambda \left( \sum_{j \in \sigma(0)^-} x_{j0} - K \right)^2,
    \label{eq:hamiltonian}
\end{align}

where $\lambda$ is a scaling parameter set by the researcher to determine how strictly the constraints must be enforced. 
The last step to convert this MTZ formulation to an Ising problem is to 
convert the cost function to an Ising Hamiltonian by performing a change of variables. 
We make the substitution
\begin{equation}
x_{ij} = \frac{1-Z_{ij}}{2} \text{, } Z_{ij} \in \{-1, 1\}
\label{eq:to_spin_variables}
\end{equation}

where $Z_{ij}$ is the Pauli-Z operator represented by the matrix 
\[
Z_{ij} = \begin{bmatrix}
1 & 0 \\
0 & -1
\end{bmatrix}.
\]
This promotes our cost function into an operator, $ \mathcal{H}_c \mapsto \hat{H}_c$, of which we can measure the expectation value on a quantum circuit.

We also implemented an amplitude encoding of the TSP as defined in \cite{stenger2024efficient}. This process takes the edges of the graph and encodes them as basis states. The resulting solution is a combination of basis states. For example, edge $\overline{AB} = \ket{00}$, $\overline{BC} = \ket{01}$, and $\overline{AC} = \ket{10}$. The solution would be a combination of each of these states, as opposed to one eigenstate which is what the QUBO formulation delivers. This process also requires a cost operator, $\hat{H_c}$, different from the one described by Equations~\ref{eq:hamiltonian} and~\ref{eq:to_spin_variables}.
\begin{figure}[t]
    \centering
    \includegraphics[width=\linewidth]{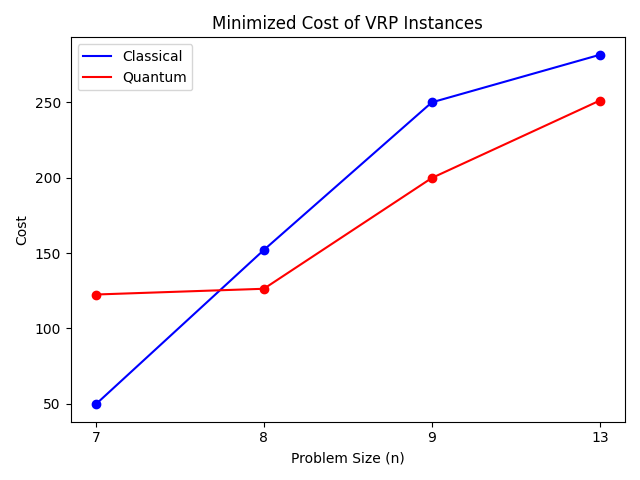}
    \caption{For the larger 8, 9, and 13 node graphs, the classical solver was unable to find the most optimal solution. 
    Our partitioning method was capable of finding more efficient solutions, with the failure for $n=7$ likely due to the graph partitioning.}
    \label{fig:Classical-Quantum Graph}
\end{figure}

\subsection{QAOA}

For each encoding our goal is to minimize the expectation value of $\hat{H}_c$. 
This is written as 
\begin{equation}
    \braket{\hat{H}_c} = \bra{\Psi(\gamma, \beta)}\hat{H}_c\ket{\Psi(\gamma, \beta)}
    \label{eqn:cost_operator_minimization}
\end{equation}
where $\ket{\Psi(\gamma, \beta)}$ is the state of the quantum computer as a function of the optimization parameters $\gamma$ and $\beta$, and $\gamma = \{\gamma_1, \gamma_2, \dots \gamma_p\}$ and $\beta = \{\beta_1, \beta_2 \dots \beta_p\}$. Once we find optimal values of $\gamma$ and $\beta$, the state, 
$\ket{\Psi(\gamma, \beta)}$, represents the solution eigenstate for the problem. An example solution eigenstate for a 3 node,
1 vehicle VRP using the QUBO method is $\ket{100110}$, which means the first, fourth, and fifth edges are traversed.

To do this, we use QAOA, which we implement with Qiskit using the noiseless Aer simulator \cite{qiskit2024aer}. 
QAOA is a hybrid variational algorithm that relies on the cost Hamiltonian and a secondary driver, or mixer, Hamiltonian. 
The hybrid classical-quantum algorithm also requires a classical optimizer to fine tune the circuit parameters.
Our choice of optimizers was the COBYLA optimizer, which stands for Constrained Optimization by Linear Approximation.

We can write the equation for the QAOA ansatz as 

\begin{equation}
    U(\gamma, \beta) = \left(\prod_{p = 0}^p e^{-i \gamma _p \hat{H}_c} e^{-i \beta _p \hat{H}_d}\right) \bigotimes_{q=0}^{N_q - 1} H_q,
    \label{eq:qaoa_state}
\end{equation}
where $\gamma$ and $\beta$ are the optimization parameters, p is the number of layers of the ansatz, $\hat{H}_d$ is the driver Hamiltonian, defined below,
$N_q$ is the number of qubits (given by Equation~\ref{eq:num_qubits}), $q$ is the $q^{th}$ qubit, and $H_q$ is the Hadamard operator on the $q^{th}$ qubit. For this paper,
we used $p = 10$ and $p = 1$.

Using Equation~\ref{eq:qaoa_state}, we can write the quantum state, $\ket{\Psi}$, as 
\begin{equation}
    \ket{\Psi} = U_p U_{p-1} \dots U_1 H^{\bigotimes N_q} \ket{0}^{\bigotimes N_q}.
\end{equation}

The driver Hamiltonian used is the standard QAOA mixer 
\begin{equation}
    \hat{H}_d = \sum_{i = 0}^{N_q - 1} X_i,
\end{equation}
where $i$ is the qubit index and $X_i$ is the Pauli-X operator on the $i^{th}$ qubit, defined as 
\[
X_i = \begin{bmatrix}
0 & 1 \\
1 & 0
\end{bmatrix}.
\]

\subsection{Circuit Level Decomposition}

After performing PLD and mapping the resulting TSPs to the quantum computer via QAOA, we perform a second, circuit level, decomposition using circuit 
cutting and knitting techniques described in \cite{tang2021cutqc} implemented with Qiskit Addon Cutting, previously called Circuit Knitting Toolbox \cite{qiskit-addon-cutting}. We performed this on select 
VRP instances as indicated in Table~\ref{tab:vrp_instances}. 
We determined which instances to cut based on the classical overhead that is created. 
Using the Qiskit automatic cut finder, we found certain VRPs, even after PLD, were too large to cut, with the classical overhead being on the order of $10^{250}$. 
To overcome this, we reduced the number of ansatz layers from ten to one. 
With the smaller circuits, we were able to reconstruct the expectation value. 
With the reduced ansatz layers, the quality of the solutions were reduced slightly.
Efforts are being made to reduce the exponential overhead created from the knitting process of CLD \cite{tang2025tensorqc}.
Our approach scales proportionally to these developments because of the ability to perform CLD on increasingly large circuits.

It is possible to reconstruct the solution eigenstate with the knitting process using methods not currently available with Qiskit \cite{tang2021cutqc}.  The Qiskit Addon Cutting does allows one to reconstruct the expectation value of the cost function, which is enough for our purposes.  Herein, we focus on the size of the subcircuits obtained from CLD and show we are able to minimize the expectation value of the cost operator, $\langle \hat{H}_c\rangle$.
We compare the depth and qubit count of the cut circuits with that of the original problem before any PLD or CLD.

\begin{figure}[t] 
    \centering
    \includegraphics[width=\linewidth]{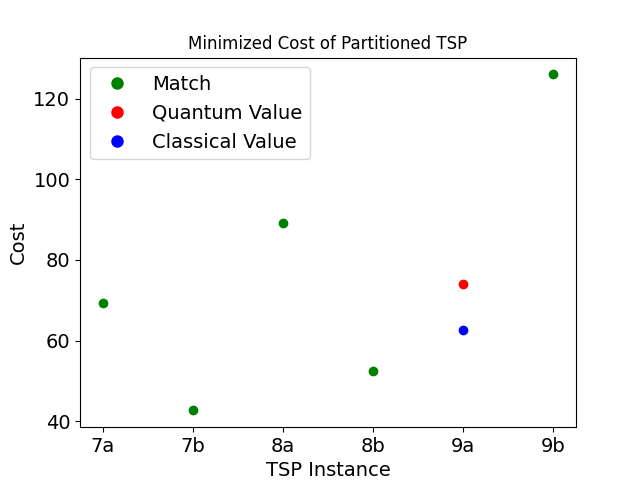}
    \caption{This graph only shows the problem level decomposition data. The initial VRP instances were partitioned into two TSP instances, "a" and "b." The green dots show the optimized cost of each TSP
    where the classical and quantum solutions
    matched. The anomaly seen at 9a is to be explored as part of future work immediately following this paper.}
    \label{fig:tsp_solutions}
\end{figure}

\section{Results}
\subsection{QUBO Encoding}
First, we present the results of the PLD, without the CLD.. Table~\ref{tab:vrp_instances} contains the exact problem sizes and whether 
or not we executed CLD. 
We performed PLD on each VRP in the table. 
We examined two seven node, 2 vehicle, VRPs and performed CLD only on one.

\begin{table}[ht]
    \centering
    \begin{tabular}{|c|c|c|}
        \hline
        n & K  & CLD\\ \hline
        7 & 2 & no \\ \hline
        7 & 2 & Yes \\ \hline
        8 & 2 & No \\ \hline
        8 & 3 & Yes\\ \hline
        9 & 2 & No \\ \hline
        9 & 3 & Yes\\ \hline
        13 & 5 & Yes\\ \hline

    \end{tabular}
    \caption{The list of VRP instances explored in this paper. CLD was only performed on the indicated problem instances. PLD was performed on each.}
    \label{tab:vrp_instances} 
    \end{table}

The results are broken into three distinct parts. (I) The partitioning of the problem: Does the partitioning of nodes for the classical solution match 
our graph partitioning? 
(II) The individual single instance VRP: Does the quantum solution to each TSP match the classical solution? 
Finally, (III): The total cost of the VRP instance. 
How does the cost obtained from our PLD solution compare to the classical cost, even if (I) and (II) differ from the classical solution. 
To explain (III) in more detail, a 6 node graph may be partitioned by METIS in a way different from the classical solution; however, if you compare the total cost of the classical and the quantum solution, they may differ by minuscule amounts making both solutions approximately equivalent.

In Fig.~\ref{fig:Classical-Quantum Graph} we present the results corresponding to the VRP instances shown in Table II.  We compare the Google-OR Tools classical method to the hybrid PLD-CLD quantum method.  Neither the classical or quantum methods converge to the optimal solution as found by brute force optimization. For the $n=7$ node case, the classical method finds a lower minimum cost than the quantum method.  However, for the $n=8$ and $n=9$ cases, the quantum method outperforms the classical method.

The second point of comparison is the individual routes of each TSP from the partitioned graph. 
We had three problems, sizes of $n = 7$, $n = 8$, and $n = 9$. 
All of these had two vehicles, meaning each VRP was divided into 2 TSP instances. 
This resulted in 6 total TSPs to be solved. 
Figure~\ref{fig:tsp_solutions} shows that of these 6 instances, 5 of them found the optimal solution. We plan to include results for the 13 node partitions in our final draft and oral presentation.
Of the 100,000 shots fired on the simulated quantum computer, each path showed up only a handful of times, with many showing up exactly once.
So while the most optimal route was found, we are looking at ways to improve the probability of measuring these states.

We performed CLD on four problem instances: $n=7$, $K = 2$; $n=8$, $K = 3$;
$n=9$, $K = 3$; and $n=13$, $K=5$.
Table~\ref{tab:pld_stats} shows the size of each problem before and after PLD, but before CLD is performed. We observe 92\% reduction in qubit count, a 92\% reduction in circuit depth, and a 98.6\% reduction in multi qubit gates for our largest problem size of 13 nodes and 5 vehicles, which was partitioned into 5 TSPs.
Performing CLD, we reduce the circuit size even more. Combined, we observe a 96\% reduction in qubit count, 95\% reduction in circuit depth, and 99.5\% reduction in multi qubit gates.
Table~\ref{tab:cld_stats_simplified} shows the further reduction from this process. CLD generates multiple subcircuits with varying sizes, we chose to include only the size of the largest subcircuit for each TSP because this represents the maximum QPU requirements for the TSP.

\begin{table}[t]
    \centering
    \begin{tabular}{|c|c|c|c|}
        \hline
        Instance & Number of Qubits & Circuit Depth & 2-Qubit Gates \\ \hline
        7.2   & 42  & 156 & 420 \\ 
        7.2a  & 12  & 45  & 48  \\
        7.2b  & 12  & 45  & 48  \\ \hline
        8.3   & 56  & 204 & 672 \\ 
        8.3a  & 6   & 18  & 12  \\
        8.3b  & 6   & 18  & 12   \\
        8.3c  & 12  & 45  & 48\\ \hline
        9.3   & 72  & 258 & 1008 \\ 
        9.3a  & 6   & 18  & 12   \\
        9.3b  & 12  & 45  & 48   \\
        9.3c  & 12  & 45  & 48   \\ \hline
        13.5  & 156 & 534 & 3432 \\ 
        13.5a & 6   & 18  & 12   \\
        13.5b & 6   & 18  & 12 \\
        13.5c & 6   & 18  & 12   \\
        13.5d & 12  & 45  & 48   \\
        13.5e & 12  & 45  & 48 \\ \hline
    \end{tabular}
    \caption{PLD-level statistics for various VRP instances and their TSP partitions, showing qubit count, circuit depth, and number of 2-qubit gates. For each entry in the first column, the first number represents the number of nodes, the second number, separated by a dot, represents the number of vehicles, and then the letters are used to distinguish partitions. 7.2 is the original seven node, two vehicle VRP, and 7.2a is the first partition.}
    \label{tab:pld_stats}
\end{table}

\begin{table}[t]
    \centering
    \begin{tabular}{|c|c|c|c|}
        \hline
        Instance & Number of Qubits & Circuit Depth & 2-Qubit Gates \\ \hline
        7.2a     & 6                & 25            & 16            \\
        7.2b     & 6                & 25            & 16            \\ \hline
        8.3a     & 3                & 10            & 4             \\
        8.3b     & 3                & 10            & 4             \\
        8.3c     & 6                & 25            & 16            \\ \hline
        9.3a     & 3                & 10            & 4             \\
        9.3b     & 6                & 25            & 16            \\
        9.3c     & 6                & 25            & 16            \\ \hline
        13.5a   & 3                & 10            & 4             \\ 
        13.5b    & 3                & 10            & 4             \\
        13.5c    & 3                & 10            & 4             \\
        13.5d    & 6                & 25            & 16            \\
        13.5e    & 6                & 25            & 16            \\ \hline
    \end{tabular}
    \caption{We show the qubit count, circuit depth and number of 2 qubit gates for the largest quantum circuit to result from CLD for each TSP instance. For each entry in the first column, the first number represents the number of nodes, the second number, separated by a dot, represents the number of vehicles, and then the letters are used to distinguish partitions. 7.2 is the original seven node, two vehicle VRP, and 7.2a is the first partition. Note that for the CLD the number after the dot is irrelevant, since each instance is a TSP containing exactly one vehicle.}
    \label{tab:cld_stats_simplified}
\end{table}
\subsection{Amplitude Encoding}
Amplitude encodings, while offering a reduction in qubit count, typically requires high circuit depth. This combined with the QAOA ansatz's high entanglement led to significantly high overhead even after reducing it as much as possible. For this reason, we were unable to perform the necessary knitting process. We obtained the results of PLD and CLD on one 10 node, 2 vehicle VRP, but are unable to reconstruct the expectation value. This impedance should be eliminated as circuit knitting processes are improved upon. We list the results of PLD on the amplitude encoding in Table~\ref{tab:amplitude_encoding_PLD} and CLD in Table~\ref{tab:amplitude_encoding_CLD}

\begin{table}[t]
    \centering

    \begin{tabular}{|c|c|c|c|}
        \hline
        Instance & Number of Qubits & Circuit Depth & 2-Qubit Gates \\ \hline
        10.2   & 8  & 1693 & 1538 \\ 
        10.2a  & 5  & 122  & 98  \\
        10.2b  & 4  & 47  & 34  \\ \hline
    \end{tabular}
    \caption{The size of the original problem and each partition before performing CLD. For each entry in the first column, the first number represents the number of nodes, the second number, separated by a dot, represents the number of vehicles, and then the letters are used to distinguish partitions. 10.2 is the original seven node, two vehicle VRP, and 10.2a is the first partition. Note that for the CLD the number after the dot is irrelevant, since each instance is a TSP containing exactly one vehicle.}
    \label{tab:amplitude_encoding_PLD}
    \end{table}
\begin{table}[t]
    \centering
    \begin{tabular}{|c|c|c|c|}
    \hline
    Instance & Number of Qubits & Circuit Depth & 2-Qubit Gates \\ \hline
    10.2a & 1 & 7 & 0 \\
    10.2b & 2 & 25 & 16 \\ \hline
    \end{tabular}
    \caption{We show the qubit count, circuit depth and number of 2 qubit gates for the largest quantum circuit to result from CLD for each TSP instance. 10.2a is the first partition. 10.2b is the second partition.}
    \label{tab:amplitude_encoding_CLD}
\end{table}
\section{Discussion}
In our PLD results, we observed that many of the potential solution eigenstates representing valid paths were measured an insignificant number of times. 
To negate this issue, we would like to explore several parts of QAOA, with the first being the driver Hamiltonian, $\hat{H}_d$, which we believe we can tailor to our 
problem to reduce optimization time as well as increase solution quality. 
We are also exploring the graph partitioning process. 
As previously mentioned, METIS is not designed to work directly with the VRP, but it may still be utilized in clever ways to increase its efficacy.
In our implementation, we did not combine the partitioning and the optimization process; it may be beneficial to couple the two so that the VRP is optimized as a whole, 
instead of just each individual TSP.

The CLD process successfully reduced the quantum resources necessary for every QUBO instance.

We observe promising results for the amplitude encoding, reducing the qubit count, depth and number of multi qubit gates significantly. We plan to leverage new circuit knitting techniques which are specifically designed to reduce unnecessary redundancy and allow for reduced classical overhead \cite{tang2025tensorqc}. As these advancements are realized, we can fully utilize our algorithm on different encodings. New knitting techniques may be designed to specifically target depth cutting more efficiently. These may prove useful because of the amplitude encoding's tendency to have lower qubit counts but higher depth. For now, it remains infeasible to implement our process with amplitude encoding.

The subcircuits created were of varying sizes. 
In other words, the difference in size between the largest and smallest subcircuits was spread out significantly, which leaves room for optimization.
We will be exploring balanced cutting so that each subcircuit is of relatively equal size. 
This will spread the quantum resources required out among all subcircuits. 
The minimum amount of quantum resources required is determined by the largest subcircuit. 
Balancing the subcircuits will further reduce the amount of quantum resources required. We also acknowledge working being done to improve the circuit knitting process such as \cite{chen2025circuit}. By implementing this approach, it may be possible to even more increase the efficiency of the cutting/knitting process.

While our work here was done entirely on Qiskit's Aer simulator, the overarching goal is to target multinode quantum computer (MNQC) architecture \cite{ang2022architectures, barral2025review}  as well as to extend our work to more optimially use classical HPCs with methods such as the Distributed QAOA \cite{kim2024distributed}.
Our preliminary results show that a hybrid PLD-CLD approach can be implemented successfully to solve problems that were previously too large to be run on today's QPUs, 
demonstrating a step toward quantum utility.

\section{Conclusion} In this work, we leverage a two level decomposition hybrid classical-quantum approach using PLD with METIS and circuit cutting with Qiskit Addon Cutting 
to successfully partition a VRP too large for quantum simulators into smaller problems that we can simulate with the Aer simulator. We observe that we can successfully 
partition the VRP using METIS to obtain satisfactory results, and we can further reduce the size of the quantum circuit representing the partitioned TSP.

We decompose a 13 node VRP, requiring 156 qubits, a 534 gate depth, and 3432 multi qubit gates into smaller circuits requiring at most 6 qubits, a depth of only 55 gates, and 16 multi qubit gates. This results in a reduction of 96\% for the qubit count, 95\% reduction in circuit depth, and 99.5\% reduction in two qubit gates.

A limitation we found is in the encoding. Due to the high entanglement of the QAOA ansatz coupled with the high depth of amplitude encoding methods, implementing the circuit knitting process with current state of the art methods is not possible.

A key challenge that we face is the customization of the graph partitioning. It is not obvious how to most efficiently partition the graph in a way that does not eliminate possible high quality solutions. The graph partitioning process utilized also does not consider the distribution of quantum resources. Our aim is to create a VRP partitioning library that is built for CLD, and optimizes the partition process in a way that evenly distributes quantum resources. This will reduce the require QPU resources for the largest circuits.  We also were met with high classical overhead for the circuit cutting process. Work is being done to 
minimize the classical overhead, making it possible to cut larger circuits. We believe that as more work is done in these areas, our approach will become increasingly more 
scalable.

This study serves as a step toward integrating quantum computing into real-world combinatorial optimization problems. Our method provides a foundation 
for further exploration of quantum-classical hybrid algorithms in logistics and supply chain management.

\section*{Acknowledgments}

This research was supported in part by the Richard T. Cheng Endowment (NC \& AM) at Old Dominion University (ODU), Naval Research Enterprise Undergraduate Summer Internship Program (AM), Office of Naval Research (ONR) Summer Faculty Research Programs (NC), and by ONR through the U.S. Naval Research Laboratory (JS \& DG). The authors thank Leo Zimianitis at CRTC and Min Dong at ITS in ODU for their help with METIS and the HPC cluster at ODU, respectively. This work was performed using computational facilities at ODU enabled by grants from the National Science Foundation (MRI grant no CNS-1828593) and Virginia's Commonwealth Technology Research Fund. Any subjective views or opinions expressed in this paper do not necessarily represent the views of the National Science Foundation or the United States Government. ChatGPT, Gemini, and Grammarly were used to improve readability; the authors reviewed and take full responsibility for the final content.

\bibliographystyle{IEEEtran}
\bibliography{circuit_cutting_thesis}

\begin{thebibliography}{10}
\providecommand{\url}[1]{#1}
\csname url@samestyle\endcsname
\providecommand{\newblock}{\relax}
\providecommand{\bibinfo}[2]{#2}
\providecommand{\BIBentrySTDinterwordspacing}{\spaceskip=0pt\relax}
\providecommand{\BIBentryALTinterwordstretchfactor}{4}
\providecommand{\BIBentryALTinterwordspacing}{\spaceskip=\fontdimen2\font plus
\BIBentryALTinterwordstretchfactor\fontdimen3\font minus \fontdimen4\font\relax}
\providecommand{\BIBforeignlanguage}[2]{{%
\expandafter\ifx\csname l@#1\endcsname\relax
\typeout{** WARNING: IEEEtran.bst: No hyphenation pattern has been}%
\typeout{** loaded for the language `#1'. Using the pattern for}%
\typeout{** the default language instead.}%
\else
\language=\csname l@#1\endcsname
\fi
#2}}
\providecommand{\BIBdecl}{\relax}
\BIBdecl

\bibitem{widuch2020}
J.~Widuch, ``Current and emerging formulations and models of real-life rich vehicle routing problems,'' in \emph{Smart Delivery Systems}.\hskip 1em plus 0.5em minus 0.4em\relax Amsterdam, The Netherlands: Elsevier, 2020, pp. 1--35.

\bibitem{ibmExonMobile}
\BIBentryALTinterwordspacing
IBM, ``Exxonmobil strives to solve complex energy challenges,'' IBM-Case-Studies, 2022. [Online]. Available: \url{https://www.ibm.com/case-studies/exxonmobil}
\BIBentrySTDinterwordspacing

\bibitem{Statista}
\BIBentryALTinterwordspacing
Statista, ``Logistics industry costs worldwide from 2010 to 2022, with forecasts until 2026,'' Statista, 2024. [Online]. Available: \url{https://www.statista.com/statistics/943500/logistics-industry-costs-worldwide/}
\BIBentrySTDinterwordspacing

\bibitem{kumar2012survey}
S.~N. Kumar and R.~Panneerselvam, ``A survey on the vehicle routing problem and its variants,'' 2012.

\bibitem{elatar2023vehicle}
S.~Elatar, K.~Abouelmehdi, and M.~E. Riffi, ``The vehicle routing problem in the last decade: variants, taxonomy and metaheuristics,'' \emph{Procedia Computer Science}, vol. 220, pp. 398--404, 2023.

\bibitem{elshaer2020taxonomic}
R.~Elshaer and H.~Awad, ``A taxonomic review of metaheuristic algorithms for solving the vehicle routing problem and its variants,'' \emph{Computers \& Industrial Engineering}, vol. 140, p. 106242, 2020.

\bibitem{irnich2014chapter}
S.~Irnich, M.~Schneider, and D.~Vigo, ``Chapter 9: Four variants of the vehicle routing problem,'' in \emph{Vehicle Routing: Problems, Methods, and Applications, Second Edition}.\hskip 1em plus 0.5em minus 0.4em\relax SIAM, 2014, pp. 241--271.

\bibitem{vidal2020concise}
T.~Vidal, G.~Laporte, and P.~Matl, ``A concise guide to existing and emerging vehicle routing problem variants,'' \emph{European Journal of Operational Research}, vol. 286, no.~2, pp. 401--416, 2020.

\bibitem{miller1960integer}
C.~E. Miller, A.~W. Tucker, and R.~A. Zemlin, ``Integer programming formulation of traveling salesman problems,'' \emph{Journal of the ACM (JACM)}, vol.~7, no.~4, pp. 326--329, 1960.

\bibitem{robinson1949hamiltonian}
J.~Robinson, \emph{On the Hamiltonian game (a traveling salesman problem)}.\hskip 1em plus 0.5em minus 0.4em\relax Rand Corporation, 1949.

\bibitem{jungnickel1999greedy}
D.~Jungnickel and D.~Jungnickel, ``The greedy algorithm,'' \emph{Graphs, networks and algorithms}, pp. 129--153, 1999.

\bibitem{gutin2002traveling}
G.~Gutin, A.~Yeo, and A.~Zverovich, ``Traveling salesman should not be greedy: domination analysis of greedy-type heuristics for the tsp,'' \emph{Discrete Applied Mathematics}, vol. 117, no. 1-3, pp. 81--86, 2002.

\bibitem{liu2023heuristics}
F.~Liu, C.~Lu, L.~Gui, Q.~Zhang, X.~Tong, and M.~Yuan, ``Heuristics for vehicle routing problem: A survey and recent advances,'' \emph{arXiv preprint arXiv:2303.04147}, 2023.

\bibitem{konstantakopoulos2022vehicle}
G.~D. Konstantakopoulos, S.~P. Gayialis, and E.~P. Kechagias, ``Vehicle routing problem and related algorithms for logistics distribution: A literature review and classification,'' \emph{Operational research}, vol.~22, no.~3, pp. 2033--2062, 2022.

\bibitem{li2007open}
F.~Li, B.~Golden, and E.~Wasil, ``The open vehicle routing problem: Algorithms, large-scale test problems, and computational results,'' \emph{Computers \& operations research}, vol.~34, no.~10, pp. 2918--2930, 2007.

\bibitem{tan2001heuristic}
K.~C. Tan, L.~H. Lee, Q.~Zhu, and K.~Ou, ``Heuristic methods for vehicle routing problem with time windows,'' \emph{Artificial intelligence in Engineering}, vol.~15, no.~3, pp. 281--295, 2001.

\bibitem{zhang2022review}
H.~Zhang, H.~Ge, J.~Yang, and Y.~Tong, ``Review of vehicle routing problems: Models, classification and solving algorithms,'' \emph{Archives of Computational Methods in Engineering}, pp. 1--27, 2022.

\bibitem{fitzek2024applying}
D.~Fitzek, T.~Ghandriz, L.~Laine, M.~Granath, and A.~F. Kockum, ``Applying quantum approximate optimization to the heterogeneous vehicle routing problem,'' \emph{Scientific Reports}, vol.~14, no.~1, p. 25415, 2024.

\bibitem{grover1996fast}
L.~K. Grover, ``A fast quantum mechanical algorithm for database search,'' in \emph{Proceedings of the twenty-eighth annual ACM symposium on Theory of computing}, 1996, pp. 212--219.

\bibitem{sato2025two}
R.~Sato, C.~Gordon, K.~Saito, H.~Kawashima, T.~Nikuni, and S.~Watabe, ``Two-step quantum search algorithm for solving traveling salesman problems,'' \emph{IEEE Transactions on Quantum Engineering}, no.~99, pp. 1--12, 2025.

\bibitem{arino2023adiabatic}
J.~F. Ari{\~n}o~Sales and R.~A. Palacios~Araos, ``Adiabatic quantum computing impact on transport optimization in the last-mile scenario,'' \emph{Frontiers in Computer Science}, vol.~5, p. 1294564, 2023.

\bibitem{kieu2019travelling}
T.~D. Kieu, ``The travelling salesman problem and adiabatic quantum computation: an algorithm,'' \emph{Quantum Information Processing}, vol.~18, no.~3, p.~90, 2019.

\bibitem{farhi2001quantum}
E.~Farhi, J.~Goldstone, S.~Gutmann, J.~Lapan, A.~Lundgren, and D.~Preda, ``A quantum adiabatic evolution algorithm applied to random instances of an np-complete problem,'' \emph{Science}, vol. 292, no. 5516, pp. 472--475, 2001.

\bibitem{tszyunsi2023quantum}
C.~Tszyunsi and I.~Beterov, ``A quantum algorithm for solving the travelling salesman problem by quantum phase estimation and quantum search,'' \emph{Journal of Experimental and Theoretical Physics}, vol. 137, no.~2, pp. 210--215, 2023.

\bibitem{srinivasan2018efficient}
K.~Srinivasan, S.~Satyajit, B.~K. Behera, and P.~K. Panigrahi, ``Efficient quantum algorithm for solving travelling salesman problem: An ibm quantum experience,'' \emph{arXiv preprint arXiv:1805.10928}, 2018.

\bibitem{martovnak2004quantum}
R.~Marto{\v{n}}{\'a}k, G.~E. Santoro, and E.~Tosatti, ``Quantum annealing of the traveling-salesman problem,'' \emph{Physical Review E—Statistical, Nonlinear, and Soft Matter Physics}, vol.~70, no.~5, p. 057701, 2004.

\bibitem{borowski2020new}
M.~Borowski, P.~Gora, K.~Karnas, M.~B{\l}ajda, K.~Kr{\'o}l, A.~Matyjasek, D.~Burczyk, M.~Szewczyk, and M.~Kutwin, ``New hybrid quantum annealing algorithms for solving vehicle routing problem,'' in \emph{International Conference on Computational Science}.\hskip 1em plus 0.5em minus 0.4em\relax Springer, 2020, pp. 546--561.

\bibitem{farhi2014quantum}
E.~Farhi, J.~Goldstone, and S.~Gutmann, ``A quantum approximate optimization algorithm,'' \emph{arXiv preprint arXiv:1411.4028}, 2014.

\bibitem{Preskill2018}
\BIBentryALTinterwordspacing
J.~Preskill, ``Quantum {C}omputing in the {NISQ} era and beyond,'' \emph{{Quantum}}, vol.~2, p.~79, Aug. 2018. [Online]. Available: \url{https://doi.org/10.22331/q-2018-08-06-79}
\BIBentrySTDinterwordspacing

\bibitem{Knill1998}
\BIBentryALTinterwordspacing
E.~Knill, R.~Laflamme, and W.~H. Zurek, ``Resilient quantum computation,'' \emph{Science}, vol. 279, no. 5349, pp. 342--345, 1998. [Online]. Available: \url{https://www.science.org/doi/abs/10.1126/science.279.5349.342}
\BIBentrySTDinterwordspacing

\bibitem{Kitaev1997}
\BIBentryALTinterwordspacing
A.~Y. Kitaev, \emph{Quantum Error Correction with Imperfect Gates}.\hskip 1em plus 0.5em minus 0.4em\relax Boston, MA: Springer US, 1997, pp. 181--188. [Online]. Available: \url{https://doi.org/10.1007/978-1-4615-5923-8_19}
\BIBentrySTDinterwordspacing

\bibitem{Shor1995}
\BIBentryALTinterwordspacing
P.~W. Shor, ``Scheme for reducing decoherence in quantum computer memory,'' \emph{Phys. Rev. A}, vol.~52, pp. R2493--R2496, Oct 1995. [Online]. Available: \url{https://link.aps.org/doi/10.1103/PhysRevA.52.R2493}
\BIBentrySTDinterwordspacing

\bibitem{tang2021cutqc}
W.~Tang, T.~Tomesh, M.~Suchara, J.~Larson, and M.~Martonosi, ``Cutqc: using small quantum computers for large quantum circuit evaluations,'' in \emph{Proceedings of the 26th ACM International conference on architectural support for programming languages and operating systems}, 2021, pp. 473--486.

\bibitem{zahedinejad2017combinatorial}
E.~Zahedinejad and A.~Zaribafiyan, ``Combinatorial optimization on gate model quantum computers: A survey,'' \emph{arXiv preprint arXiv:1708.05294}, 2017.

\bibitem{stenger2024efficient}
J.~Stenger, S.~Crowe, J.~Diaz, R.~Rodriguez, D.~Gunlycke, and J.~Ptasinski, ``Efficient encoding of the traveling salesperson problem on a quantum computer,'' 2024.

\bibitem{VehicleRoutingTextBook}
D.~V. Paolo~Toth, \emph{VEHICLE ROUTING Problems, Methods, and Applications}, K.~Scheinberg, Ed.\hskip 1em plus 0.5em minus 0.4em\relax Mathematical Optimization Society, Society for Industrial and Applied Mathematics, 2015.

\bibitem{METIS}
\BIBentryALTinterwordspacing
Karypis, ``Metis,'' Github, 2023. [Online]. Available: \url{https://github.com/KarypisLab/METIS}
\BIBentrySTDinterwordspacing

\bibitem{chrisochoides1989automatic}
N.~P. Chrisochoides, C.~E. Houstis, E.~N. Houstis, S.~Kortesis, and J.~R. Rice, ``Automatic load balanced paritioning strategies for pde computations,'' in \emph{Proceedings of the 3rd International Conference on Supercomputing}, 1989, pp. 99--107.

\bibitem{chicano2025combinatorial}
F.~Chicano, G.~Luque, Z.~A. Dahi, and R.~Gil-Merino, ``Combinatorial optimization with quantum computers,'' \emph{Engineering Optimization}, pp. 1--26, 2025.

\bibitem{goldsmith2024beyond}
D.~Goldsmith and J.~Day-Evans, ``Beyond qubo and hobo formulations, solving the travelling salesman problem on a quantum boson sampler,'' \emph{arXiv preprint arXiv:2406.14252}, 2024.

\bibitem{glos2022space}
A.~Glos, A.~Krawiec, and Z.~Zimbor{\'a}s, ``Space-efficient binary optimization for variational quantum computing,'' \emph{npj Quantum Information}, vol.~8, no.~1, p.~39, 2022.

\bibitem{qiskit2024aer}
A.~Javadi-Abhari, M.~Treinish, K.~Krsulich, C.~J. Wood, J.~Lishman, J.~Gacon, S.~Martiel, P.~D. Nation, L.~S. Bishop, A.~W. Cross, B.~R. Johnson, and J.~M. Gambetta, ``Quantum computing with {Q}iskit,'' 2024.

\bibitem{qiskit-addon-cutting}
A.~M. Bra\'{n}czyk, A.~{Carrera Vazquez}, D.~J. Egger, B.~Fuller, J.~Gacon, J.~R. Garrison, J.~R. Glick, C.~Johnson, S.~Joshi, E.~Pednault, C.~D. Pemmaraju, P.~Rivero, I.~Shehzad, and S.~Woerner, ``{Qiskit addon: circuit cutting},'' \url{https://github.com/Qiskit/qiskit-addon-cutting}, 2024.

\bibitem{tang2025tensorqc}
W.~Tang and M.~Martonosi, ``Tensorqc: Towards scalable distributed quantum computing via tensor networks,'' \emph{arXiv preprint arXiv:2502.03445}, 2025.

\bibitem{chen2025circuit}
X.~Chen, Z.~Chen, P.~Zhu, X.~Cheng, and Z.~Guan, ``Circuit partitioning and transmission cost optimization in distributed quantum circuits,'' \emph{IEEE Transactions on Computer-Aided Design of Integrated Circuits and Systems}, 2025.

\bibitem{ang2022architectures}
J.~Ang, G.~Carini, Y.~Chen, I.~Chuang, M.~A. DeMarco, S.~E. Economou, A.~Eickbusch, A.~Faraon, K.-M. Fu, S.~M. Girvin \emph{et~al.}, ``Architectures for multinode superconducting quantum computers.(2022),'' \emph{arXiv preprint arXiv:2212.06167}, 2022.

\bibitem{barral2025review}
D.~Barral, F.~J. Cardama, G.~D{\'\i}az-Camacho, D.~Fa{\'\i}lde, I.~F. Llovo, M.~Mussa-Juane, J.~V{\'a}zquez-P{\'e}rez, J.~Villasuso, C.~Pi{\~n}eiro, N.~Costas \emph{et~al.}, ``Review of distributed quantum computing: from single qpu to high performance quantum computing,'' \emph{Computer Science Review}, vol.~57, p. 100747, 2025.

\bibitem{kim2024distributed}
S.~Kim, T.~Luo, E.~Lee, and I.-S. Suh, ``Distributed quantum approximate optimization algorithm on integrated high-performance computing and quantum computing systems for large-scale optimization,'' \emph{arXiv preprint arXiv:2407.20212}, 2024.

\end{thebibliography}
\end{document}